\documentclass[11pt,reqno]{amsart}

\usepackage{amsmath,amssymb,amsthm,bm,mathrsfs,braket,sasaki,color,graphicx}
\newtheorem{thm}{Theorem}[section]

\theoremstyle{definition}

\theoremstyle{remark}
\newtheorem{rem}[thm]{Remark}

\usepackage{color}

\title[A Short Remark on the Polaron]{A Short Remark on the Polaron in the Semi-relativistic Pauli-Fierz Model}
\author{Itaru Sasaki}
\thanks{I. S.'s work was partly supported by Research supported by KAKENHI Y22740087, and was performed through the Program for Dissemination of Tenure-Track System funded by the Ministry of Education and Science, Japan}
\address{Fiber-Nanotech Young Researcher Empowerment Center, Shinshu University,
Asahi 3--1--1, Matsumoto 390--8621, Japan.}
\email{isasaki@shinshu-u.ac.jp}

\date{\today}



\begin{document}
%%%%%%%%%%%%%%%%
\maketitle

\begin{abstract}
We consider the polaron of the spinless semi-relativistic Pauli-Fierz model. 
The Hamiltonian of the model is defined by $H(\bP) = \sqrt{(\bP-d\Gamma(\bk)+e\bA)^2+M^2}+d\Gamma(\omega_m)$,
where $\bP\in\BR^3$ is the momentum of the polaron, $d\Gamma(\cdot)$ denotes the second quantization operator
and $\omega_m=|\bk|+m$ denotes the dispersion relation of the photon with virtual mass $m\geq 0$.
Let $E(\bP)$ be the lowest energy of $H(\bP)$. In this paper, we prove the inequality
\begin{align*}
  E(\bP-\bk) - E(\bP) + \omega_m(\bk) \geq m,
\end{align*}
for all $\bP, \bk\in\BR^3$. 
\end{abstract}

%%%%%%%%%%%%%%%%%%%%%%%%%%%%%%%%%%%%%%%%%%%%

\setlength{\baselineskip}{15pt}

\section{Definition and Main Result}
We consider a system of a charged spinless relativistic particle interacting with the quantized radiation field with fixed total momentum 
$\bP\in\BR^3$.
The Hilbert space of the model is the Fock space
\begin{align}
  \cF :=  \bigoplus_{n=0}^\infty \left[ \bigotimes_\mathrm{sym}^n L^2(\BR^3\times \{1,2\})\right]
\end{align}
with $\tensor_\mathrm{sym}^0 L^2(\BR^3\times\{1,2\}) =:\BC$.
Let $d\Gamma(\cdot)$ be the second quantization operator.
 The Hamiltonian of the model is defined by
\begin{align}
  H(\bP) := \sqrt{(\bP-d\Gamma(\bk) + e\bA)^2 + M^2} + d\Gamma(\ome_m),
\end{align}
where $e$ is the coupling constant, $M>0$ is the mass of the particle, $\ome_m(\bk)=|\bk|+m$ is the photon dispersion relation
 and $\bA$ is the quantized magnetic vector potential at the origin $\bx=0$.
See \cite{Miyao-Spohn:2009} for more detailed definition.
The self-adjointness of $H(\bP)$ was studied in \cite{Miyao-Spohn:2009} and \cite[Corollary 7.62]{LHB11}.
We assume the following
%%%%%%%%%%%%%%%%
\begin{itemize}
 \item[{[H.1]}] There exist a dense domain $\cD\subset \cF$ such that, for all $\bP\in\BR^3$, 
\begin{align}
 &\cD \subset \dom((\bP-d\Gamma(\bk)+e\bA)^2) \cap \dom(d\Gamma(\ome_m))
\end{align}
and $H(\bP)$ is essentially self-adjoint on $\cD$.
\end{itemize}
%%%%%%%%%%%%%%%%
Clearly, $H(\bP)$ is bounded from below, and we define the ground state energy:
\begin{align}
  E(\bP) := \inf \spec(H(\bP))
\end{align}
The main theorem of this paper is the following:
%%%%%%%%%%%%%%%% main theorem
\begin{thm}{\label{mthm}}
 For all $m\geq 0$, the inequality
\begin{align}
 E(\bP-\bk) - E(\bP) +\ome_m(\bk) \geq m, \quad \bk,\bP \in\BR^3 \label{gap}
\end{align}
holds.
\end{thm}
%%%%%%%%%%%%%%%%%
 \begin{rem}
We define
\begin{align}
  \Delta(\bP) := \inf_{\bk\in\BR^3} \{E(\bP-\bk) - E(\bP) +\ome_m(\bk) \}
\end{align}
As discussed in \cite{Miyao-Spohn:2009}, $\Delta(\bP)$ is the spectral gap at the lowest energy of $H(\bP)$.
The inequality \eqref{gap} implies that the spectral gap is open uniformly in $\bP$.
To establish \eqref{gap}, our photon dispersion relation $|\bk|+m$ was important. 
Similar result may not holds for the standard massive dispersion relation $\ome_m(\bk)=\sqrt{\bk^2+m^2}$.
 \end{rem}
 \begin{rem}
This is the remarkable that \eqref{gap} holds for all $\bP$ and there is no restriction on the other parameters.
This fact is different from the non-relativistic QED where the spectral gap is open only for $\bP^2/2M<1$. 
The uniform spectral gap is a nature of the relativistic dynamics.
 \end{rem}
 \begin{rem}
To prove \eqref{gap}, we only use the operator monotonicity. 
Unfortunately, our method works only for the spinless case.
 \end{rem}
 \begin{rem}
It is strongly expected that $H(\bP)$ with $m=0$ has ground state for all $\bP\in\BR^3$ and $e\in\BR$ 
under suitable conditions including the infrared regularization. 
By \eqref{gap}, the massive Hamiltonian $H(\bP), (m>0)$ has a ground state for all $m>0$.
But, in our opinion, it is difficult to construct a ground state of massless model as the massless limit $m\downarrow 0$.
 \end{rem}

\section{Proof of Theorem \ref{mthm}}
By the variational principle, \eqref{gap} follows from the operator inequality
\begin{align}
 H(\bP-\bk) + \ome_m(\bk) \geq m + H(\bP), \label{1}
\end{align}
on $\cD$. We set
\begin{align}
  K(\bP) :=   \sqrt{(\bP-d\Gamma(\bk)+e\bA)^2+M^2}.
\end{align}
Then \eqref{1} is equivalent to 
\begin{align}
  K(\bP-\bk)  \geq K(\bP)- |\bk|,  \label{2}
\end{align}
on $\cD$. By L\"{o}wner-Heinz inequality, \eqref{2} follows form
\begin{align}
  K(\bP-\bk)^2 \geq (K(\bP)-|\bk|)^2, \label{3}
\end{align}
in the sense of the quadratic forms on $\cD$. 
By expanding the square of the operator, we know that \eqref{3} is equivalent to
\begin{align}
 |\bk| \cdot K(\bP) \geq \bk \cdot (\bP-d\Gamma(\bk)+e\bA). \label{4}
\end{align}
By the L\"owner Heinz inequality, \eqref{4} follows from 
\begin{align}
 \bk^2 K(\bp)^2 \geq [ \bk \cdot (\bP-d\Gamma(\bk)+e\bA) ]^2 \label{5}
\end{align}
We set $\mathbf{B}=(B_1,B_2,B_3):=\bP-d\Gamma(\bk)+e\bA$.
Note that, for all self-adjoint operators $a,b$, it holds that $ab+ba\leq a^2+b^2$
in the sense of the quadratic forms on $\dom(a)\cap\dom(b)$.
Then we have
\begin{align}
 \left( \sum_{j=1}^3 k_j B_j \right)^2
 &= \sum_{j,l=1}^3 k_j B_j k_l B_l \\
 & \leq \frac{1}{2}\sum_{j,l} \left(k_j^2 B_j^2 + k_l^2 B_l^2 \right) \\
 & = \sum_{j} k_j^2 \sum_{l} B_l^2 = |\bk|^2 \cdot \mathbf{B}^2 \\
 & \leq |\bk|^2 \cdot (\mathbf{B}^2+M^2)
\end{align}
on $\cD$. Therefore \eqref{5} holds.

%\bibliographystyle{amsplain}
%\bibliography{sasaki-biblio}

\providecommand{\bysame}{\leavevmode\hbox to3em{\hrulefill}\thinspace}
\providecommand{\MR}{\relax\ifhmode\unskip\space\fi MR }
\providecommand{\MRhref}[2]{%
  \href{http://www.ams.org/mathscinet-getitem?mr=#1}{#2}
}
\providecommand{\href}[2]{#2}

\end{document}